\documentclass[twocolumn,showpacs,aps,prl,superscriptaddress]{revtex4-1}

\usepackage{graphicx}
\usepackage{bm}
\usepackage{amsmath,amssymb,amsfonts}
\usepackage{hyperref}
\usepackage{color}
\usepackage{subfigure}
\usepackage{xr}
\usepackage{pdfpages}

\begin{document}

\title{Anomalous Crystal Symmetry Fractionalization on the Surface of Topological Crystalline Insulators}
\author{Yang Qi}
\affiliation{Institute for Advanced Study, Tsinghua University,
  Beijing 100084, People's Republic of China}
\affiliation{Perimeter Institute for Theoretical Physics, Waterloo, Ontario
  N2L 2Y5, Canada} 
\author{Liang Fu}
\affiliation{Department of Physics, Massachusetts Institute of
  Technology, Cambridge, Massachusetts 02139, USA}

\begin{abstract}
  The surface of a three-dimensional topological electron system often
  hosts symmetry-protected gapless surface states. With the effect of
  electron interactions, these surface states can be gapped out
  without symmetry breaking by a surface topological order, in which
  the anyon excitations carry anomalous symmetry fractionalization
  that cannot be realized in a genuine two-dimensional system. We show
  that for a mirror-symmetry-protected topological crystalline
  insulator with mirror Chern number $n=4$, its surface can be gapped
  out by an anomalous $\mathbb Z_2$ topological order, where all
  anyons carry mirror-symmetry fractionalization $M^2=-1$. The
  identification of such anomalous crystalline symmetry
  fractionalization implies that in a two-dimensional $\mathbb Z_2$
  spin liquid the vison excitation cannot carry $M^2=-1$ if the spinon
  carries $M^2=-1$ or a half-integer spin.
\end{abstract}


\maketitle

The advent of topological insulators (TIs)~\cite{Hasan2010,
  QiZhang2011, Moore2010} and topological superconductors
(TSCs)~\cite{Schnyder2008} has greatly broadened our understanding of
topological phases in quantum systems. While the concepts of TIs and
TSCs originate from topological band theory of noninteracting
electrons or quasiparticles, recent theoretical
breakthroughs~\cite{FidkowskiKitaev2011, Ryu2012, Yao2013, Qi2DZ82013,
  FidkowskiPRX2013, Wang2014a} have found that interactions can, in
principle, change fundamental properties of these topological phases
dramatically, thus creating a new dimension to explore.
In particular, interactions can drive the gapless Dirac fermion
surface states of three-dimensional (3D) TIs and TSCs into
topologically ordered phases that are gapped and
symmetry preserving. Nonetheless, such a surface manifests the
topological property of the bulk in a subtle but unambiguous way: its
anyon excitations have anomalous symmetry transformation properties,
which cannot be realized in any two-dimensional (2D) system with the
same symmetry.

Given the profound consequences of interactions in TIs and TSCs, the
effect of interactions in topological phases protected by spatial
symmetries of crystalline solids, commonly referred to as topological
crystalline insulators (TCIs)~\cite{Fu2011a}, is now gaining wide
attention.  A wide array of TCI phases with various crystal symmetries
have been found in the framework of topological band
theory~\cite{AndoFu2015, BernevigReview}. One class of TCIs has been
predicted and observed in the IV-VI semiconductors SnTe,
Pb$_{1-x}$Sn$_x$Se and Pb$_{1-x}$Sn$_x$Te~\cite{HsiehSnTe2012,
  AndoSnTe2012, PolandPbSnSe2012, HasanPbSnTe2012}. The topological
nature of these materials is warranted by a particular mirror symmetry
of the underlying rocksalt crystal, and is manifested by the presence
of topological surface states on mirror-symmetric crystal
faces. Remarkably, there surface states were found to become gapped
under structural distortions that break the mirror
symmetry~\cite{vidya1,vidya2}, confirming the mechanism of crystalline
protection unique to TCIs~\cite{HsiehSnTe2012}.

The study of interacting TCIs has just begun. A recent work by Isobe
and Fu~\cite{IsobeFu2015} shows that in the presence of interactions,
the classification of 3D TCIs protected by mirror symmetry (i.e., the
SnTe class) reduces from being characterized by an integer known as
the mirror Chern number~\cite{Teo2008} (hereafter denoted by $n$) to
its $\mathbb Z_8$ subgroup. This implies that interactions can turn
the $n=8$ surface states, which consists of eight copies of 2D massless
Dirac fermions with the same chirality, into a completely trivial
phase that is gapped, mirror symmetric and without intrinsic
topological order.  It remains an open question what interactions can
do to TCIs with $n \neq 0 \mod
8$. 
In this work, we
take the first step to study strongly interacting TCI
surface states 
for the case $n=4 \mod 8$.

Our main result is that the surface of a 3D TCI with mirror Chern
number $n=4\mod8$ can become a gapped and mirror-symmetric state with
$\mathbb Z_2$ topological order. Remarkably, the mirror symmetry acts
on this state in an anomalous way that all three types of anyons carry
fractionalized mirror quantum number $\tilde M^2=-1$ (in this Letter we
use $\tilde M$ to represent the projective representation of mirror
symmetry $M$ acting on an anyon), which cannot be realized in a purely
2D system. Furthermore, the anomalous mirror-symmetry
fractionalization protects a twofold degeneracy between two
mirror-symmetry-related edges.
Such anomalous mirror-symmetry fractionalization cannot be realized in
a 2D system, including a 2D $\mathbb Z_2$ spin liquid
state~\cite{wen1991,*wen1991a}. Hence, our finding constrains the
possible ways of fractionalizing the mirror symmetry in a 2D
$\mathbb Z_2$ spin liquid~\cite{wenpsg,Essin2013}.  Brief reviews of
3D TCIs, 2D $\mathbb Z_2$ spin liquids, and their edge theory are
available in the Supplemental Material~\footnote{See the Supplemental
  Material, which includes Refs.~\cite{LuBFU,kitaev,sstri,Tanaka2005},
  for reviews of 3D TCIs, 2D $\mathbb Z_2$ spin liquids, and their
  edge theory.}.

\emph{Noninteracting TCIs.}
We begin by considering noninteracting TCIs protected by the mirror
symmetry $x \rightarrow -x$. With the mirror symmetry, the extra U(1)
symmetry in a TCI does not change the classification in 3D compared
to a mirror-protected topological crystalline superconductor. Hence,
for convenience, we choose a TCI as our starting point, although the
U(1) symmetry plays no role in this work. As we will explain in
Sec.~\ref{sec:mz2} of the Supplemental Material, in order to produce
an anomalous $\mathbb Z_2$ surface topological order, the mirror
operation must be defined as a $\mathbb Z_2$ symmetry with the
property $M^2=1$.  In our previous works on spin-orbit coupled
systems, the mirror operation $M'$ acts on electron's spin in addition
to its spatial coordinate, which leads to $M'^2= -1$.  Nonetheless,
one can redefine the mirror operation by combining $M'$ with the U(1)
symmetry of charge conservation $c \rightarrow i c$, which restores
the property $M^2=1$. We note that without the U(1) symmetry only $M$
satisfying $M^2=+1$ protects nontrivial topological crystalline
superconductors.


The mirror TCIs are classified by the mirror Chern number $n$, defined
for single-particle states on the mirror-symmetric plane $k_x=0$ in
the 3D Brillouin zone.  The states with mirror eigenvalues $1$ and
$-1$ form two different subspaces, each of which has a Chern number
denoted by $n_+$ and $n_-$ respectively. This leads to two independent
topological invariants for noninteracting systems with mirror
symmetry: the total Chern number $n_T = n_+ + n_-$ and the mirror
Chern number $n = n_+ - n_-$.

The TCI with a nontrivial mirror Chern number $n$ has gapless surface
states consisting of $n$ copies of massless Dirac fermions, described
by the following surface Hamiltonian
\begin{equation}
  \label{hs}
  H_s = v \sum_{A=1}^{n}
  \psi^\dagger_A (k_x \sigma_y - k_y \sigma_x ) \psi_A,
\end{equation}
where the two-dimensional fermion fields $\psi_A(x,y)$ transform as
the following under mirror operation:
\begin{equation}
  \label{eq:Mpsi}
  M:  \psi_A (x, y) \rightarrow \sigma_x \psi_A(-x, y).
\end{equation}
The presence of mirror symmetry~\eqref{eq:Mpsi} forbids any Dirac mass
term $\psi^\dagger_A \sigma_z \psi_B$. As a result, the surface states
described by Eq.~\eqref{hs} cannot be gapped by fermion bilinear terms,
for any flavor number $n$.
 
We emphasize that the above Dirac fermions on the surface of a 3D TCI
cannot be realized in any 2D system with mirror symmetry, as expected
for symmetry-protected topological phases in general.
According to the Hamiltonian (\ref{hs}), the surface states with
$k_x=0$ within a given mirror subspace are chiral as they all move in
the same direction~\cite{HsiehSnTe2012}.  In contrast, in any 2D
system single-particle states within a mirror subspace cannot be
chiral (this is demonstrated with a 2D lattice model in
Sec.~\ref{sec:piflux} of the Supplemental Material~\cite{Note1}).

\emph{$U(1)$ Higgs phase and $\mathbb Z_2$ topological order.}
In this work, we study interacting surface states of TCIs with
$n=4$. Starting from four copies of Dirac fermions in the
noninteracting limit, we will introduce microscopic interactions and
explicitly construct a $\mathbb Z_2$ topologically-ordered phase on the TCI
surface, which is gapped and mirror symmetric.

Our construction is inspired by the work of \citet{Senthil2006} and
\citet{SenthilMotrunich} on fractionalized insulators. We construct on
the surface of an $n=4$ TCI a Higgs phase with an $xy$-order parameter
$\langle b\rangle\neq0$, which is odd under the mirror symmetry and
gaps the Dirac fermions.  Next, we couple these gapped fermions to
additional degrees of freedom $a_\mu$ that are introduced to mimic a
$U(1)$ gauge field.  This gauge field $a_\mu$ plays three crucial
roles: 
(i) the coupling between matter and $a_\mu$ restores the otherwise
broken $U(1)$ symmetry and, thus, the mirror symmetry along with it;
(ii) the Goldstone mode is eaten by the gauge boson and becomes
massive; (iii) since the $xy$-order parameter carries $U(1)$ charge 2,
the $U(1)$ gauge group is broken to the $\mathbb{Z}_2$ subgroup in the
Higgs phase. Because of these properties, the Higgs phase thus
constructed is a gapped and mirror-symmetric phase with $\mathbb{Z}_2$
topological order.

We now elaborate on the construction (details of this construction can
be found in Sec.~\ref{sec:intr-u1z2} of the Supplemental
Material~\cite{Note1}). First, we relabel the fermion flavors
$A=1,\ldots, 4$ using a spin index $s=\uparrow, \downarrow$ and a
$U(1)$-charge index $a=\pm$ (unrelated to the electric charge).  We
take fermion interactions that are invariant under both the $SU(2)$
spin rotation and the $U(1)$ rotation
\begin{equation}
  \label{eq:u1psi}
  U(1): \psi_{as} \rightarrow
  e^{i a \theta} \psi_{as}, \; a=\pm
\end{equation}
Moreover, we introduce a boson field $b(x,y)$ that carries
$U(1)$-charge 2 and is odd under mirror symmetry,
\begin{equation}
  \label{eq:u1m}
    U(1): b\rightarrow  e^{i2\theta}b,\quad
    M: b(x,y) \rightarrow - b(-x,y),
\end{equation}
and couple this boson to the
massless Dirac fermions as follows
\begin{equation}
  \label{eq:hbf}
  H_{bf}= V b^\dagger \psi^\dagger_{a s} \tau^-_{ab} \sigma_z \psi_{bs}
  + \text{H.c.}
\end{equation}
When these bosons condense, $\langle b\rangle\neq0$ spontaneously
breaks both the $U(1)$ and mirror symmetry, and gappes out the
fermions.

Finally, we introduce another boson vector field $a_\mu(x,y)$, which
couples to $b$ and $\psi_{as}$ through minimal coupling. An effective theory
of this system has the following form,
\begin{equation}
  \label{eq:Lsfd}
  \begin{split}
    \mathcal L = &-i\psi^\dagger_{s}\alpha^\mu(\partial_\mu+ia_\mu
    \tau_z)\psi_s + (b \psi^\dagger_s \tau^+ \sigma_z
    \psi_s + \text{H.c.})\\
    &+\frac{1}{2g} |(\partial_\mu - 2i a_\mu)b|^2 + r |b|^2
    + u |b|^4 + F_{\mu\nu} F^{\mu\nu},
  \end{split}
\end{equation}
where the matrices $\alpha^0=1$, $\alpha^x=\sigma_y$ and
$\alpha^y=\sigma_x$. Furthermore, we add to the effective action an
interation term $UN^2$, where
$N=\psi^\dagger\tau_z\psi+2b^\dagger b-\nabla\cdot\bm E$
($E_i=F_{0i}=\partial_0a_i-\partial_ia_0$ is the electric field
strength). In the limit of $U\rightarrow\infty$, this enforces the
local constraint $N=0$.  As a result, the bare fermion $\psi_s$ and
boson $b$ are no longer low-energy excitations, since adding them to
the ground state violates the constraint $N=0$ and costs an energy
$U$. Therefore, in the low-energy effective model $\psi_s$ and $b$ must
be screened by the gauge field $a_\mu$ and become quasiparticles
$\tilde\psi_{as}=\psi_{as}e^{ia\theta}$ and $\tilde{b}=be^{2i\theta}$,
where the operator $e^{in\theta}$ creates $n$ gauge charge of $a_\mu$
and restores the constraint $N=0$. In terms of these quasiparticles,
the effective theory becomes
\begin{equation}
  \label{eq:Lsfd2}
  \begin{split}
    \mathcal L =
    &-i\tilde\psi^\dagger_{s}\alpha^\mu(\partial_\mu+ia_\mu
    \tau_z)\tilde\psi_s + (\tilde b\tilde\psi^\dagger_s \tau^+ \sigma_z
    \tilde\psi_s + \text{H.c.})\\
    &+\frac{1}{2g} |(\partial_\mu - 2i a_\mu)\tilde b|^2 + r |\tilde
    b|^2 + u |\tilde b|^4 + F_{\mu\nu} F^{\mu\nu}.
  \end{split}
\end{equation}
Furthermore, a $U(1)$ gauge symmetry emerges in the low-energy Hilbert
space defined by the local constraint $N=0$~\footnote{Here the
  constraint $N=0$ can be enforced locally because the U(1) symmetry
  defined in Eqs.~\eqref{eq:u1psi} and \eqref{eq:u1m} is an on-site
  symmetry operation. Particularly the U(1) rotation of the Dirac
  fermion $\psi$ is a local rotation between four flavors of Dirac
  fermions, which arise from the four flavors of Dirac fermions in the
  bulk of the 3D TCI, as explained in Sec.~\ref{sec:mirr-topol-cryst}
  of the Supplemental Material~\cite{Note1}.}. Specifically, the
constraint is the Gauss law and it restricts the low-energy Hilbert
space to states that are invariant under the gauge transformation
\begin{equation}
  \label{eq:Utheta}
  U_\phi:\tilde\psi_s\rightarrow e^{i\phi\tau_z}\tilde\psi_s,\quad
  \tilde b\rightarrow \tilde be^{2i\phi},\quad
  a_\mu\rightarrow a_\mu-\partial_\mu\phi.
\end{equation}

In this effective theory with the emergent $U(1)$ gauge field,
condensing the boson $\tilde b$ no longer breaks the global $U(1)$ and
the mirror symmetries, as it instead breaks the $U(1)$ gauge symmetry to
$\mathbb Z_2$. Naively, the mirror symmetry maps
$\langle\tilde b\rangle$ to $-\langle\tilde b\rangle$. However these
two symmetry breaking vacuua are equivalent because they are related
by the gauge symmetry transformation
$U_{\pi/2}:\tilde b\rightarrow -\tilde b$. This restoration of mirror
symmetry becomes clearly manifested if we assume that $\tilde b$ and
$\tilde\psi_s$ transform projectively under mirror symmetry with the
additional U(1) gauge transformation $U_{\pi/2}$,
\begin{equation}
  \label{eq:chisu2}
  \tilde M: \tilde\psi_s(r)\rightarrow
  i\tau_z\otimes\sigma_x\tilde\psi_s(r^\prime);
  \tilde b(r)\rightarrow\tilde b(r^\prime).
\end{equation}

This Higgs phase obtained by condensing charge-2 $\tilde b$ field
indeed has a $\mathbb Z_2$ topological order when the number of Dirac
fermions is $n=4$~\cite{Wang2014a,Metlitski2014}. This can be
understood by identifying the Bogoliubov quasiparticle $\tilde \psi$
and vortices as the anyons $e$, $m$, and $\epsilon$ (see
Sec.~\ref{sec:mz2} of the Supplemental Material~\cite{Note1} for the
definition of the notation) in the $\mathbb Z_2$ topological
order. $\tilde\psi$ becomes the $\epsilon$ anyon as both are
fermions. As the Higgs field gaps out four Dirac fermions, there are
four Majorana fermions, or two complex fermion zero modes, in each
vortex core. Hence, there are two types of vortices whose core has
even or odd fermion parity, respectively. In the case of $n=4$, it can
be shown that the vortices carry Bose statistics (see
Sec.~\ref{sec:topol-order-higgs} of the Supplemental
Material~\cite{Note1} for details), and they are mapped to the $m$ and
$e$ anyons in the $\mathbb Z_2$ topological order, respectively.

\emph{Mirror-symmetry fractionalization.}
Now we consider how the mirror symmetry acts in the $\mathbb Z_2$ spin
liquid phase described by Eq.~\eqref{eq:Lsfd2}. In this effective
theory, the $\tilde\psi$ field is the fermionic anyon
$\epsilon$. Equation~\eqref{eq:chisu2} implies that it carries
$\tilde M^2=-1$.

\begin{figure}[tb]
  \centering
  \includegraphics{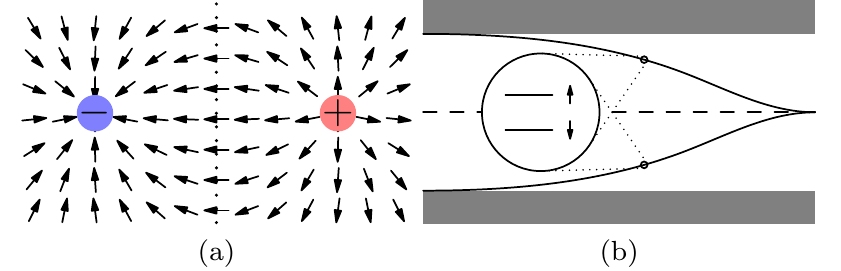}
  \caption{(a) A vortex and an antivortex. The direction of the arrow
    represents the phase of the Higgs field $\langle\tilde b\rangle$.
    The dotted line is the mirror axis. (b) Illustration of the fermion
    spectrum flow from the left to the right as we create a
    vortex-antivortex pair from the vacuum and move them far apart. In
    this process, four vortex core states are separated from the bulk
    spectrum, two from the conducting band and two from the valence
    band, and become degenerate zero modes. As illustrated in the
    inset, the core states are twofold degenerate with spin $s=\pm1$.}
  \label{fig:vortices}
\end{figure}

Next, we consider how the mirror symmetry acts on the $m$ anyon, which
is a vortex of the Higgs field where all core states are empty. Since
the mirror symmetry preserves the Higgs field $\langle\tilde b\rangle$
but maps $x$ to $-x$, it maps a vortex to an antivortex. Therefore we
consider a mirror-symmetric configuration with one vortex and one
antivortex, as shown in Fig.~\ref{fig:vortices}(a). The symmetry
fractionalization of $\tilde M^2=\pm1$ can be detected from the $M$
parity of the fermion wave function with such a vortice-antivortex
pair~\cite{QiCSF}: for two bosonic vortices, the mirror parity is equal to
$\tilde M^2$.

From Eq.~\eqref{eq:Lsfd2}, we get the fermion Hamiltonian
\begin{equation}
  \label{eq:Hchi}
  H=v\tilde\psi_s^\dagger\left(k_x\sigma_y-k_y\sigma_x\right)\tilde\psi_s
  +\tilde\psi_s^\dagger(\langle\tilde b\rangle\tau^+
  +\langle\tilde b\rangle^\ast\tau^-)\sigma_z\tilde\psi_s.
\end{equation}
It has a particle-hole symmetry
$\Xi:\tilde\psi\rightarrow\sigma_z\tilde\psi$ which maps $H$ to
$\Xi H\Xi=-H$. This implies that its spectrum is
symmetric with respect to zero. Assume that
the dimension of the whole Hilbert space is $4N$; there are
$2N$ states with positive energy and $2N$ states with negative
energy. For the vortex configuration in Fig.~\ref{fig:vortices}(a), there
are four complex zero modes, two from each vortex core, which are all
unoccupied. Therefore, excluding these four states there are $2N-2$
states with negative energy, which are all occupied in the fermion
wave function.

Next, we consider the mirror eigenvalues of these $2N-2$ occupied
fermion states. Since $\tilde\psi$ carries $\tilde M^2=-1$, each state
has mirror eigenvalue $\lambda_M=\pm i$. Because both $H$, $\Xi$ and
$M$ are diagonal in pseudospin $s=\uparrow,\downarrow$, all occupied
states are pseudospin doublets, and two states in each doublet have
the same $\lambda_M$.  Hence, the mirror eigenvalue of all occupied
states, organized as $N-1$ doublets, is $(-1)^{N-1}=-1$. Therefore, the
wave function of two empty vortices is odd under mirror symmetry,
which implies that the $e$ particle has the symmetry fractionalization
$\tilde M^2=-1$.  Combining the results that both $e$ and $\epsilon$
carry $\tilde M^2=-1$, we conclude that the $m$ anyon also has
$\tilde M^2=-1$ (see the discussion in Sec.~\ref{sec:mz2} of the
Supplemental Material~\cite{Note1}).

In summary, the gapped $\mathbb Z_2$ surface state we constructed has
an anomalous mirror-symmetry fractionalization that both types of
anyons carry $\tilde M^2=-1$, which cannot be realized in a genuine 2D
system. For comparison, it is shown in Sec.~\ref{sec:mz2} of the
Supplemental Material~\cite{Note1} that applying the same construction
to a 2D lattice model results in a $\mathbb Z_2$ phase with a
different mirror-symmetry fractionalization which is not anomalous.

\begin{figure}[htbp]
  \centering
  \includegraphics{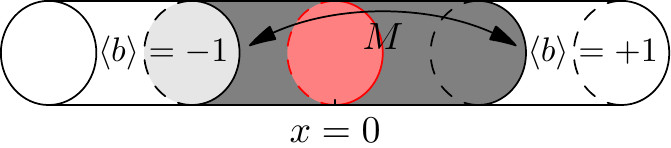}
  \caption{Two mirror-symmetric edges of a $\mathbb Z_2$ surface
    topological order. The mirror symmetry maps $x$ to $-x$ with
    respect to the mirror plane marked by the red disk at $x=0$. The
    surface topological order marked by the shade terminates at edges
    against two ordered regions with $\langle b\rangle=\pm1$,
    respectively.}
  \label{fig:subway}
\end{figure}

\emph{Mirror anomaly.}
The anomalous crystal symmetry fractionalization presented in the
surface topological order implies a symmetry-protected topological
degeneracy associated with the edges of the surface topological
ordered region. This mirror anomaly is a remnant of the anomalous
surface fermion modes in the free-fermion limit. To see this, we
consider the setup presented in Fig.~\ref{fig:subway}, in which the
$\mathbb Z_2$ surface topologically ordered state is terminated at two
edges symmetric with respect to the mirror plane, by two regions with
opposite $\langle b\rangle=\pm1$ on either side of the mirror plane,
respectively.

This setup itself does not break the mirror symmetry, and all local
excitations can be gapped everywhere on the surface. In particular,
since the $\mathbb Z_2$ topological order is not chiral, its edge can
be gapped out by condensing either $e$ or $m$ anyons on the
edge~\cite{Barkeshli2014a}. The edges next to an ordered phase with
$\langle b\rangle\neq0$ are $e$ edges, as condensing $e$ breaks the
global U(1) symmetry. A $\mathbb Z_2$ spin liquid state on an infinite
cylinder has four degenerate ground states $|\Psi_a\rangle$, each has
one type of anyon flux $a=1,e,m,\epsilon$ going through the
cylinder. On a finite cylinder with two $e$ edges, only
$|\Psi_1\rangle$ and $|\Psi_e\rangle$ remain degenerate, because
adding an $m$ or $\epsilon$ anyon on the edge costs a finite
energy. In a generic $\mathbb Z_2$ state, this degeneracy can be
further lifted by tunneling an $e$ anyon between the two edges,
$H_t=\lambda e_L^\dagger e_R^\dagger+\text{H.c.}$, where
$e_{L,R}^\dagger$ creates two $e$ anyons on the two edges,
respectively. However, the $e$ anyon carries $\tilde M^2=-1$;
therefore, the tunneling term $H_t$ is odd under mirror and, thus,
forbidden by $M$. As a result, this twofold topological degeneracy is
protected by the mirror symmetry even in the limit of
$L\rightarrow0$. This argument is formulated using the effective edge
Lagrangian in the Supplemental Material~\cite{Note1}.

In the limit of $L\rightarrow0$, this topological degeneracy becomes a
local degeneracy protected by the mirror symmetry. Therefore if the
$\mathbb Z_2$ topological order is killed by collapsing two gapped
edges, the ground state is either gapless or mirror-symmetry breaking,
and this cannot be avoided regardless of edge types because all types
of anyons have $\tilde M^2=-1$. This topological degeneracy reveals
the anomalous nature of this mirror-symmetry
fractionalization. Furthermore, if we collapse two gapless edges of
the $\mathbb Z_2$ state, the edges remain gapless because the anyon
tunneling is forbidden by $M$. Hence we get a gapless domain wall with
central charge $c=1+1=2$, which recovers the edge with four chiral
fermion modes in the aforementioned free-fermion limit. This is
explained in more detail in Sec.~\ref{sec:edge-lutt-theory} of the
Supplemental Material~\cite{Note1}.

\emph{Conclusion.}
In this Letter, we show that the surface of a 3D mirror TCI with
mirror Chern number $n=4$, containing four gapless Dirac fermion modes
in the free limit, can be gapped out without breaking the mirror
symmetry by a $\mathbb Z_2$ topological order. This surface
$\mathbb Z_2$ topological order has an anomalous mirror-symmetry
fractionalization in which all three types of anyons carry
fractionalized mirror-symmetry quantum number $\tilde M^2=-1$, and
such a topological order cannot be realized in a purely 2D system.

Our finding also puts constraints on possible ways to fractionalize
the mirror symmetry in a 2D $\mathbb Z_2$ quantum spin
liquid~\cite{Essin2013,BarkeshliX}. The result of this work indicates
that the combination that both the $e$ and $m$ carry the
fractionalized $\tilde M^2=-1$ is anomalous and cannot be realized in a
2D $\mathbb Z_2$ spin liquid. Furthermore, our result can be easily
generalized to also rule out the combination that the $e$ anyon
carries spin-$\frac12$ and $m$ anyon carries
$\tilde M^2=-1$~\cite{HermeleAspen}, because if $e$ carries
$\tilde M^2=+1$, we can define a new mirror symmetry
$M^\prime=Me^{i\pi S^z}$, for which both $e$ and $m$ carry
$(\tilde M^\prime)^2=-1$ and, therefore, this combination is also
anomalous. In summary, our finding implies that the vison must carry
$\tilde M^2=+1$ in a $\mathbb Z_2$ spin liquid where the spinon
carries a half-integer spin.

\begin{acknowledgments}
  We thank Chen Fang and Senthil Todadri for invaluable discussions.
  Y.Q. is supported by National Basic Research Program of China
  through Grant No. 2011CBA00108 and by NSFC Grant
  No. 11104154. L.F. is supported by the DOE Office of Basic Energy
  Sciences, Division of Materials Sciences and Engineering, under
  Award No. DE-SC0010526. This research was supported in part by
  Perimeter Institute for Theoretical Physics. Research at Perimeter
  Institute is supported by the Government of Canada through Industry
  Canada and by the Province of Ontario through the Ministry of
  Research and Innovation.
\end{acknowledgments}

\bibliography{TCI,additional}

\clearpage


\section*{Supplementary material}

# Supplemental Material

Yang Qi[1,2] and Liang Fu[3]

[1]*Institute for Advanced Study, Tsinghua University, Beijing 100084, China*
[2]*Perimeter Institute for Theoretical Physics, Waterloo, ON N2L 2Y5, Canada*
[3]*Department of Physics, Massachusetts Institute of Technology, Cambridge, MA 02139, USA*

## I. MIRROR TOPOLOGICAL CRYSTALLINE INSULATOR

In this section we briefly review non-interacting TCIs protected by the mirror symmetry $x \to -x$. An example of such a TCI with mirror Chern number $n$ can be constructed from the following low-energy Dirac fermion Hamiltonian in three dimensions

$$H = \sum_{A=1}^{n} c_A^\dagger (v \sum_{i=1}^{3} k_i \Gamma_i + m\Gamma_0) c_A \qquad (1)$$

where the $4 \times 4$ matrices $\Gamma_i$ with $i = 0, 1, 2, 3, 4$ form a set of anticommuting Dirac Gamma matrices, and $a$ is the flavor index. The fermion field $c_A$ transforms as follows under mirror:

$$M : c_A(x, y, z) \to \Gamma_{14} c_A(-x, y, z), \quad \Gamma_{14} \equiv i\Gamma_1 \Gamma_4. \qquad (2)$$

It is straightforward to check that

$$H = MHM^{-1}, \qquad (3)$$

i.e., the Hamiltonian (1) is invariant under mirror. As we will explain in Section II, for the purpose of this work, it is conceptually convenient to regard the four-components of the Dirac fermion in (1) as unrelated to electron's spin. Relatedly, the mirror operation defined here is a $Z_2$ symmetry with the property $M^2 = 1$. In our previous works on spin-orbit coupled systems, the mirror operation $M'$ acts on electron's spin in addition to its spatial coordinate, which leads to $M'^2 = -1$. Nonetheless, one can redefine the mirror operation by combining $M'$ with the $U(1)$ symmetry of charge conservation $c \to ic$, which restores the property $M^2 = 1$.

As discussed in the main text, the mirror Chern number $n$ is defined as the difference $n = n_+ - n_-$, where $n_\pm$ are the the Chern numbers of single-particle states with mirror eigenvalue $M = \pm 1$ on the mirror-symmetric plane $k_x = 0$ in the 3D Brillouin zone. Starting

from a trivial insulator with $n_+ = n_- = 0$, a TCI with a nonzero mirror Chern number $n$ is obtained by going through a band inversion transition that corresponds to reversing the sign of the Dirac mass $m$ for all $n$ flavors in the Hamiltonian (1). The total Chern number is unchanged during this transition.

## II. MIRROR SYMMETRY FRACTIONALIZATION IN $\mathbb{Z}_2$ SPIN LIQUID STATE

In this section we discuss the possible ways of mirror symmetry fractionalization in a two-dimensional $\mathbb{Z}_2$ spin liquid state. We first discuss this in the context of a bosonic system, then generalize the results to systems with fermions.

The $\mathbb{Z}_2$ topological order contains four types of anyons: $1$, $e$, $m$, and $\epsilon$, denoting the trivial bosonic particle, the bosonic spinon, the vison, and the fermion spinon, respectively. Except the trivial anyon 1, the other three anyons can carry fractionalized mirror symmetry representation $\tilde{M}^2 = \pm 1$. Because the $\epsilon$ anyon is a bound state of $e$ and $m$ anyons, its mirror symmetry fractionalization can be derived from the ones of $e$ and $m$ anyons as

$$\tilde{M}_\epsilon^2 = -\tilde{M}_e^2 \tilde{M}_m^2, \tag{4}$$

where the extra minus sign appears because the $e$ and $m$ anyons have a nontrivial mutual statistics, which gives rise to the Fermi statistics of the $\epsilon$ anyon [1, 2]. Using this relation we can enumerate the four possible ways of mirror symmetry fractionalization in a $\mathbb{Z}_2$ state, as listed in Table I.

TABLE I. Four different ways of mirror symmetry fractionalization.

| No. | $e$ | $m$ | $\epsilon$ | Realization |
|-----|-----|-----|------------|-------------|
| 1   | $+1$ | $+1$ | $-1$ | Standard toric code |
| 2   | $-1$ | $+1$ | $+1$ | Some spin liquids |
| 3   | $+1$ | $-1$ | $+1$ | Same as above |
| 4   | $-1$ | $-1$ | $-1$ | Surface of 3D TCI |

Among the four states listed in Table I, three can be realized in a purely two-dimensional system. In the first state the $e$ and $m$ anyons transform trivially under mirror symmetry, and this is realized in the standard Kitaev toric code model [3]. In the second state, the $e$

anyon has $\tilde M^2 = -1$, and this is realized in many mean field spin liquid states, for example the $Q_1 = Q_2$ state on the kagome lattice [4]. Without additional global symmetries the $e$ and $m$ anyons are interchangable in a $\mathbb{Z}_2$ topological order, so the third state is equivalent to the second state.

Since the first three states can all be realized in purely 2D systems, this leaves the fourth state as the only possibility of a $\mathbb{Z}_2$ state with an anomalous mirror symmetry fractionalization. Indeed, as we see in the main text, this state can only be realized on the surface of a 3D nontrivial TCI.

In the above discussion we considered a purely bosonic system where all physical degrees of freedom are bosons. However, the system we study in the main context is an interacting fermionic system. In this case, besides the aforementioned anyons $\{1, e, m, \epsilon\}$ the system also has local fermion $f$, which carries no gauge charge or flux but obeys Fermi statistics, and it can form bound states with other types of anyons. Particularly the bound state $\epsilon \times f$ is a boson and has a $-1$ mutual statistics with either $e$ or $m$ anyon. Therefore it is similar to the $e$ or $m$ anyon, and the set of anyons $\{1, e, \epsilon \times f, m \times f\}$ have the same self and mutual statistics as the original set of $\{1, e, m, \epsilon\}$. Hence if we neglect the fermion parity, the three anyons $e$, $m$ and $\epsilon \times f$ becomes indistinguishable.

The mirror symmetry fractionalization of $\epsilon \times f$ can be derived from the fractionalization of $\epsilon$ anyon and the mirror symmetry representation of $f$ fermion,

$$\tilde M^2_{\epsilon \times f} = \tilde M^2_\epsilon M^2_f = -\tilde M^2_e \tilde M^2_m M^2_f. \tag{5}$$

Hence for the anomalous mirror symmetry fractionalization we are considering (State 4 in Table I), $\tilde M^2_{\epsilon \times f} = -M^2_f$. If the mirror symmetry acts as $M^2 = -1$ on $f$ fermion, the bound state $\epsilon \times f$ transforms trivially under $M$ and can be condensed without breaking $M$. Such an anyon condensation confines all anyons and drives the surface to a trivial state. Therefore the surface topological order is only anomalous if fermion carries $M^2_f = +1$, which is consistent with the fact that $M^2_f = -1$ does not protect topological crystalline superconductors in 3D (for TCI the mirror symmetry can be redefined to have $M^2 = +1$ because of the charge U(1) symmetry, as explained in the main text.) This is the reason why we choose to define $M^2 = +1$ for the fermions in this work. Furthermore, in the case of $M^2_f = +1$, in the anomalous case all the three bosonic anyons $e$, $m$ and $\epsilon \times f$ has $\tilde M^2 = -1$. Therefore we do not need to examine the fermion parity of the anyon excitations when studying their mirror

symmetry fractionalization.

### III. CONSTRUCTION OF A SURFACE $\mathbb{Z}_2$ TOPOLOGICAL ORDER.

In this section, we review the detailed steps of obtaining a $\mathbb{Z}_2$ spin liquid state in Eq. (7) of the main text from the surface effective theory in Eq. (6) of the main text.

The effective theory in Eq. (6), containing the surface Dirac fermions $\psi_{as}$, the scalar boson field $b$ and the U(1) vector field $a_\mu$, only has a global U(1) symmetry. Specifically, the effective Lagrangian is invariant under the U(1) symmetry transformation where $a_\mu$ is invariant the $\psi_{as}$ and $b$ transforms as in Eqs. (3) and (4) of the main text, respectively. We emphasize that in this U(1) symmetry transformation the phase factor $e^{i\phi}$ is a space-time-independent constant.

Next, we promote this U(1) global symmetry to a U(1) gauge symmetry by introducing an interaction term $UN^2$,

$$\mathcal{L} = -i\psi_s^\dagger \alpha^\mu(\partial_\mu + ia_\mu \tau_z)\psi_s + (b\psi_s^\dagger \tau^+ \sigma_z \psi_s + h.c.) + \frac{1}{2g}|(\partial_\mu - 2ia_\mu)b|^2 + r|b|^2 + u|b|^4 + F_{\mu\nu}F^{\mu\nu} + UN^2, \tag{6}$$

where $N$ is the U(1) charge density $N = \psi^\dagger \tau^z \psi + 2b^\dagger b - \nabla \cdot \boldsymbol{E}$. Here $\boldsymbol{E}$ is the field strength of the $a_\mu$ vector field $E_i = F_{0i} = \partial_0 a_i - \partial_i a_0$. In the limit of strong interaction: $U \to +\infty$, the last term in Eq. (6) dominates. At the leading order of the perturbation theory, we only consider the $U$ term, and the system has degenerate ground states described by field configurations satisfying the constraint $N = 0$ everywhere. Higher orders of perturbation theory can then be described as an effective field theory with the constraint $N = 0$.

The local constraint $N = 0$ can be understood as an emergent gauge symmetry. First, this constraint is the same as the Gauss law,

$$\nabla \cdot \boldsymbol{E} = \psi^\dagger \tau^z \psi + 2b^\dagger b. \tag{7}$$

Second, under the U(1) gauge transformation

$$U(1)_\phi : \psi_s \to e^{i\phi\tau_z}\psi_s, \quad b \to be^{2i\phi}, \quad a_\mu \to a_\mu - \partial_\mu \phi, \tag{8}$$

the phase of the local Hilbert space changes by $e^{iN\phi}$. Therefore, the low-energy Hilbert space satisfying the constraint of $N = 0$ consists of configurations that are invariant under the U(1) gauge transformation in Eq. (8). In other words, enforcing the constraint of $N = 0$

in the large-$U$ limit restricts the low-energy dynamics of the model to an effective Hilbert space with an emergent U(1) gauge symmetry.

Another way of understanding this emergent U(1) gauge symmetry is to notice that the bare excitations $\psi_s^\dagger$ and $b^\dagger$ violates the local constraint $N = 0$, and creates excitations with U(1) charge one and two, respectively. Because of the $U$-term, these excitations carry interaction energy $U$ and $4U$, respectively, and are therefore excluded from the low-energy Hilbert space in the large-$U$ limit. In other words, a fermion excitation $\psi_s^\dagger$ must be screened by gauge field fluctuations and form a quasiparticle excitation $\tilde{\psi}_s^\dagger = \psi_s^\dagger e^{i\theta}$ that satisfies the constraint of $N = 0$, where the operator $e^{i\theta}$ creates a unit of gauge charge: $[e^{i\theta(x)}, \nabla \cdot \boldsymbol{E}(x')] = \delta^2(x - x')$ and restores the constraint in Eq. (7). Similarly a boson excitation $b^\dagger$ must be screened by $e^{2i\theta}$ and become a quasiparticle excitation $\tilde{b}^\dagger = b^\dagger e^{2i\theta}$. These fermionic and bosonic quasiparticle excitations create excited states within the low-energy Hilbert space defined by the constraint of $N = 0$, and obey the gauge transformation in Eq. (8) and Eq. (8) of the main text. They are used to construct the low-energy effective theory in Eq. (7) of the main text.

Starting from the low-energy effective theory with the emergent U(1) gauge symmetry, condensing the bosonic quasiparticle $\tilde{b}$ gives rise to a Higgs phase with no mirror-symmetry-breaking, no gapless excitations and a $\mathbb{Z}_2$ topological order, in contrast to a different Higgs phase obtained by condensing the bare boson $b$ in the original free-fermion model, where the mirror symmetry is broken. In the main text, it is explained that the $b$ field is odd under mirror reflection [see Eq. (4) of the main text]. Hence, a condensate with $\langle b \rangle \neq 0$ is mapped to a different condensate with $-\langle b \rangle$ under the mirror symmetry, indicating that the Higgs phase of $\langle b \rangle \neq 0$ breaks the mirror symmetry. However, the condensate of $\langle \tilde{b} \rangle$ does not break the mirror symmetry. Although naively mirror symmetry maps $\tilde{b}$ to $-\tilde{b}$, these two field configurations are related by a gauge transformation of $U_{\pi/2}$ that maps $-\tilde{b}$ back to $\tilde{b}$, and they represent the same physical state in the low-energy Hilbert space defined by the constraint $N = 0$, as a result of the emergent U(1) gauge symmetry. Therefore the condensate of $\langle \tilde{b} \rangle \neq 0$ nolonger breaks the mirror symmetry. This is more clearly manifested by the redefinition of the mirror symmetry action in Eq. (9) of the main text.

Moreover, the condensation of $\tilde{b}$ gaps out all excitations and creates a $\mathbb{Z}_2$ topological order. On one hand, the condensation generates a mass gap for the fermionic quasiparticle $\tilde{\psi}$ through the Yukawa coupling $\left( \tilde{b} \tilde{\psi}_s^\dagger \tau^+ \sigma_z \tilde{\psi}_s + \text{h.c.} \right)$. On the other hand, it also generates a

mass gap for the gauge boson through the Higgs mechanism. Since the Higgs field $\tilde{b}$ carries two units of the gauge charge [see Eq. (8) of the main text], it breaks the emergent U(1) gauge symmetry to a $\mathbb{Z}_2$ gauge symmetry. This $\mathbb{Z}_2$ gauge symmetry is generated by the remaining gauge transformation $U_\pi$, which maps $\tilde{\psi}_s$ to $-\tilde{\psi}_s$. Therefore, in the Higgs phase, the fermionic quasiparticle $\tilde{\psi}_s$ carries a unit of $\mathbb{Z}_2$ gauge charge, and it couples to a $\mathbb{Z}_2$ gauge field, which is the reminant of the emergent U(1) gauge field. As shown in Sec. IV, this Higgs phase has the $\mathbb{Z}_2$ topological order.

## IV. TOPOLOGICAL ORDER IN THE HIGGS PHASE

In this section, we show that the Higgs phase obtained in the main text after condensing $\langle \tilde{b} \rangle$ has the $\mathbb{Z}_2$ topological order. Naively the condensation of the Higgs field breaks the U(1) gauge symmetry to $\mathbb{Z}_2$ and therefore the remaining gauge fluctuation is described by a $\mathbb{Z}_2$ gauge theory. However, integrating out the gapped Dirac fermions can produce nontrivial Berry phases that alters the statistics of the vison excitations in the $\mathbb{Z}_2$ gauge theory. Here we show that in our case the visons still have trivial bosonic statistics and the Higgs phase has the $\mathbb{Z}_2$ topological order.

After the Higgs condensation the Bogoliubov quasiparticles and the vortices become anyon excitations. First, the Bogoliubov quasiparticle $\tilde{\psi}$ is a fermion and carries a $\mathbb{Z}_2$ charge. Therefore it can be identified as the $\epsilon$ anyon. Next, we consider the vortices of the $\tilde{b}$ condensate in this Higgs phase. We note that such vortices host four Majorana zero-energy core states in the vortex core, one from each flavour of Dirac fermion, and these four Majorana zero modes can be grouped into two complex fermion zero modes. Hence there are four degenerate core states, labeled by the occupation of these two complex zero modes $|n_1 n_2\rangle$. This four-fold degeneracy can be lifted by fermion interaction, after which the two states with even fermion parity ($|00\rangle$ and $|11\rangle$) and the two states with odd fermion parity ($|01\rangle$ and $|01\rangle$) are still degenerate. We first consider the brading statistics of the empty vortex $|00\rangle$. To do that we define an antiunitary transformation

$$\mathcal{T}: \tilde{\psi}_s \to i\tau^z \otimes i\sigma^y \tilde{\psi}_s; \tilde{b} \to \tilde{b}, \tag{9}$$

which can be viewed as an analog of the time-reversal symmetry. It is easy to check that the effective Lagrangian of the Dirac fermions with $s=\uparrow$ and $s=\downarrow$ in Eq. (7) is invariant under

$\mathcal{T}$ separately. This implies that their contribution to the Berry phase of exchanging two empty vortices must be $\pm 1$. Therefore combining the contributions from $s = \uparrow\downarrow$ the total exchange phase is $(\pm 1)^2 = +1$, indicating that the empty vortex is a boson. Moreover, it carries a $\pi$ flux of the emergent U(1) gauge field (because $\tilde{b}$ carries charge-2), and therefore has a $-1$ mutual statistics with the $\tilde{\psi}$ fermion. Hence the empty vortex can be identified as the $e$ anyon. Furthermore, the half-filled vortex with one fermion mode occupied can be viewed as the bound state of an empty vortex and a Bogoliubov quasiparticle $\tilde{\psi}$, and therefore corresponds to the $m$ anyon. In conclusion, in the case of $n = 4$ the Higgs phase has the $\mathbb{Z}_2$ topological order, and this is in contrast to other more exotic topological orders obtained in the case when $n$ is not a multiple of four [5, 6].

## V. MIRROR SYMMETRY IN THE 2D SQUARE LATTICE $\pi$-FLUX MODEL.

If mirror symmetry is not broken, a 3D TCI discussed in this work has an anomalous surface that either has gapless Dirac fermions or has a topological order in which anyons carry anomalous mirror symmetry fractionalization. It is worthwhile to compare these anomalous TCI surface states with the $\pi$-flux model on the square lattice. This genuine 2D model has Dirac fermions that can be gapped out without breaking mirror symmetry or adding interactions. Furthermore, if a $\mathbb{Z}_2$ topologically ordered state is constructed using the same approach as described in the main text, the resulting state has mirror symmetry fractionalization that is not anomalous.

The $\pi$-flux model on a 2D square lattice is known to host 2D Dirac fermions at half-filling. By linearizing the tight-binding model and keeping track of the mirror symmetry, we find [7, 8] the following Dirac Hamiltonian for the $\pi$-flux model

$$H_{2D} = v \sum_{a=1,2} \sum_{s=\uparrow,\downarrow} \psi_{as}^\dagger (k_x \sigma_y - (-1)^a k_y \sigma_x) \psi_{as}. \tag{10}$$

where the combination of $a = 1, 2$ and $\sigma_z = \pm 1$ denotes four lattice sites in a unit cell, and $s$ labels electron's spin [9]. Each fermion field $\psi_{as}$ transforms under the mirror symmetry of the square lattice as described by Eq. (2). Comparing Eq. (10) to Eq. (1) in the main text, we find that unlike the case of TCI surface states, states on the 2D square lattice are non-chiral within each mirror sector, as can be seen from the sign factor $(-1)^a$. In this case,

a coupling between the two Dirac points

$$\psi^\dagger_{1s}\sigma_x\psi_{2s} + h.c. \qquad (11)$$

opens up a gap, while preserving the mirror symmetry.

This construction of $\mathbb{Z}_2$ spin liquid states also applies to the 2D Dirac fermion effective model in Eq. (10) derived from the $\pi$-flux model. However the mirror symmetry fractionalization in the constructed $\mathbb{Z}_2$ spin is different. Because the Dirac fermions can now be gapped out by the coupling in Eq. (11), we modify the way the $b$ boson couples to the Dirac fermion in Eq. (5) to the following form,

$$H'_{bf} = V b^\dagger \psi^\dagger_{as} \tau^-_{ab} \sigma_x \psi_{bs} + \text{h.c.}. \qquad (12)$$

Hence in the $\mathbb{Z}_2$ spin liquid state the anyons $\tilde{b}$ and $\tilde{\psi}$ transforms trivially without an extra U(1) gauge transformation under mirror symmetry, and this implies that the $\tilde{\psi}$ (or the $\epsilon$) anyon carries trivial mirror symmetry fractionalization $\tilde{M}^2 = +1$. Therefore only one of the $e$ and $m$ anyons carries $\tilde{M}^2 = -1$ and the other one still carries $\tilde{M}^2 = +1$. This crystal symmetry fractionalization is allowed in a purely two-dimensional system, as discussed in Sec. I.

## VI. EDGE LUTTINGER THEORY

In this section we derive the conclusions on edges of a $\mathbb{Z}_2$ spin liquid state using the effective field theory of the edges, which has the form of a Luttinger theory.

The edge of a $\mathbb{Z}_2$ spin liquid state is described by the following single-channel Luttinger Liquid effective action

$$\mathcal{L}_{\text{edge}} = \frac{1}{\pi}\partial_t\phi\partial_y\theta - \frac{u}{2\pi}\left[K(\partial_y\theta)^2 + K^{-1}(\partial_y\phi)^2\right]. \qquad (13)$$

where $\theta$ and $\phi$ describes edge fluctuations of $e$ anyon and $m$ anyon flux, respectively, and they obey the commutation relation that $[\phi(y), \theta(y')] = i\frac{\pi}{2}\text{sgn}(y - y')$. In a $\mathbb{Z}_2$ state both two $e$ or two $m$ particles fuse into a trivial excitation and therefore the perturbations $e^{\pm 2i\theta}$ and $e^{\pm 2i\phi}$ are allowed on the edge, and in general the gap will be gapped out by either one of the two terms, unless the gaplessness is protected by some other global symmetries. (In our setup, each edge by itself is not symmetric under the mirror symmetry but mapped to the

other edge instead; hence the mirror symmetry puts no constraints on these terms.) Here in our setup, the $\mathbb{Z}_2$ phase is terminated by the ordered phase $\langle b \rangle \neq 0$, where $b \sim e^{2i\theta}$. This corresponds to an edge that is gapped by the perturbation $e^{\pm 2i\theta}$, which is described by the following action,

$$\mathcal{L}_{\text{edge}} = \frac{1}{\pi}\partial_t\phi\partial_y\theta - \frac{u}{2\pi}\left[K(\partial_y\theta)^2 + K^{-1}(\partial_y\phi)^2\right] + \lambda\cos 2\theta. \tag{14}$$

In the limit that $\lambda \to \infty$, the $\theta$ field condenses at the minima $\theta = 0, \pi$, which is consistent with the order state $\langle b \rangle = +1$. Similarly for the right edge, the potential term is $-\lambda\cos 2\theta$, giving rise to minima at $\theta = \frac{\pi}{2}, \frac{3\pi}{2}$ as the ordered state on the right has $\langle b \rangle = -1$. Putting the left and the right edges together, the effective Lagrangian is the following,

$$\mathcal{L} = \sum_{a=L,R}\left\{\frac{1}{\pi}\partial_t\phi_a\partial_y\theta_a - \frac{u}{2\pi}\left[K(\partial_y\theta_a)^2 + K^{-1}(\partial_y\phi_a)^2\right]\right\} \\ + \lambda(\cos 2\theta_L - \cos 2\theta_R). \tag{15}$$

The above Hamiltonian is invariant under the mirror transformation if we assume that $\theta$ field transforms under mirror as

$$M : \theta_{L,R} \to \theta_{R,L} + \frac{\pi}{2}. \tag{16}$$

Since the $e$ anyon is identified as $e \sim e^{i\theta}$, this is consistent with the mirror symmetry fractionalization $\tilde{M}^2 e = -e$.

In the effective action of Eq. (15) each $\theta_a$ has two degenerate minima, so there are four degenerate minimal configurations labeled by $|\theta_L\theta_R\rangle$, where $\theta_L = 0, \pi$ and $\theta_R = \frac{\pi}{2}, \frac{3\pi}{2}$. However the states $|\theta_L\theta_R\rangle$ and $|\theta_L + \pi, \theta_R + \pi\rangle$ are related by the $\mathbb{Z}_2$ gauge transformation and therefore represent the same physical state. Therefore there is a two-fold ground state degeneracy labeled by $\Delta\theta = \theta_R - \theta_L = \pm\frac{\pi}{2}$.

When we bring the two edges together, in a normal $\mathbb{Z}_2$ state there will be tunneling terms that couple the two edges together and lift the two-fold degeneracy by locking the relative phase of $\Delta\theta$. To the lowest order such tunneling term has the following form,

$$\mathcal{L}_t = \tau_1 e^{i\theta_R - i\theta_L} + \tau_2 e^{i\theta_R + i\theta_L} + \text{c. c.}. \tag{17}$$

Using the mirror symmetry transformation in Eq. (16) we see that the mirror invariance requires that $\tau_1 = \tau_1^*$ and $\tau_2 = 0$, so the tunneling term takes the form $\mathcal{L}_t = \tau_1\cos(\Delta\theta)$,

which does not lift the degeneracy between $\Delta\theta = \pm\frac{\pi}{2}$. Therefore this two-fold ground state degeneracy remain exact as we collapse the two gapped edges and eliminate the $\mathbb{Z}_2$ topological order, and this can only be realized at the surface of a nontrivial 3D TCI.

Furthermore, if we collapse two gapless edges we can recover the gapless Dirac fermion modes in the free fermion limit. To study gapless edges, we can fine tune the parameters in the effective action of Eq. (15) such that the edges are gapless. Without the mirror symmetry, the two gapless edges will be gapped by the tunneling term $\tau_2 e^{i\theta_R+i\theta_L}$ in Eq. (17) when we collapse the two edges. However, the mirror symmetry enforces that $\tau_2 = 0$, which implies that the two edges remain gapless (the $\tau_1$ term becomes a constant in this limit since $\theta_R \to \theta_L$ as $L \to 0$). Therefore when we collapse the two edges, the $\mathbb{Z}_2$ spin liquid region becomes a domain wall separating two gapped regions, and this domain wall hosts gapless modes with central charge $c = 1 + 1 = 2$, where the left and right edges each contribute $c = 1$. This is consistent with the conclusion that in the free fermion limit such a domain wall hosts four Dirac fermion modes [10].



\end{document}